\documentclass[lettersize,journal]{IEEEtran}
\usepackage{amsmath,amsfonts}
\usepackage{algorithmic}
\usepackage{algorithm}
\usepackage{array}
\usepackage[caption=false,font=normalsize,labelfont=sf,textfont=sf]{subfig}
\usepackage{textcomp}
\usepackage{stfloats}
\usepackage{url}
\usepackage{verbatim}
\usepackage{graphicx}
\usepackage{cite}
\usepackage{amssymb}
\usepackage{booktabs, tabularx}
\usepackage{hyperref}
\usepackage{units}
\usepackage{multirow,array}
\usepackage{balance}
\usepackage{soul}
\usepackage{color}
\usepackage{cleveref}
\usepackage{wrapfig}
\usepackage{booktabs}
\usepackage{pifont}
\usepackage[utf8]{inputenc}
\usepackage{makecell, tablefootnote, longtable}

\newcommand{\eg}{e.\,g.,\ }
\newcommand{\ie}{i.\,e.,\ }

\usepackage{soul}
\usepackage{xcolor}
\definecolor{newgreen}{RGB}{34,139,34}

\newcommand{\re}[1] {{\color{black} #1}}
\newcommand{\zr}[1] {{\color{black} #1}}
\newcommand{\yc}[1] {{\color{black} #1}}

\newcommand{\sstitle}[1]{\smallskip\noindent\textbf{#1.\/}}

\newcommand{\cmark}{\ding{51}}
\newcommand{\xmark}{\ding{55}}
\begin{document}

\title{A Comprehensive Survey on Heart Sound Analysis in the Deep Learning Era}

\author{Zhao Ren, University of Bremen, Germany\\
Yi Chang, Imperial College London, United Kingdom\\
Thanh Tam Nguyen, Griffith University, Australia\\
Yang Tan, Beijing Institute of Technology, China\\
Kun Qian, Beijing Institute of Technology, China\\
Björn W. Schuller, University of Augsburg, Germany \& CHI, MRI, Technical University of Munich, Germany \& GLAM, Imperial College London, United Kingdom
\thanks{\emph{Corresponding author}: Zhao Ren (zren@uni-bremen.de).
This work has been submitted to the IEEE for possible publication. Copyright may be transferred without notice, after which this version may no longer be accessible.}
%\thanks{}
%\thanks{Y.\ Chang and B.\,W.\ Schuller are with the Imperial College London, United Kingdom (y.chang20@imperial.ac.uk, schuller@ieee.org).}
%\thanks{T.\,T.\ Nguyen is with the Griffith University, Australia (t.nguyen19@griffith.edu.au).}
%\thanks{Y.\ Tan and K.\ Qian are with the Beijing Institute of Technology, China (qian@bit.edu.cn).}
%\thanks{B.\,W.\ Schuller is also with the University of Augsburg, Germany.}
% \thanks{% this sentence is for arxiv version only.
% This work has been submitted to the IEEE for possible publication. Copyright may be transferred without notice, after which this version may no longer be accessible.}
}

%\author{IEEE Publication Technology,~\IEEEmembership{Staff,~IEEE,}
        % <-this % stops a space
%\thanks{This paper was produced by the IEEE Publication Technology Group. They are in Piscataway, NJ.}% <-this % stops a space
%\thanks{Manuscript received April 19, 2021; revised August 16, 2021.}}

\markboth{Journal of \LaTeX\ Class Files,~Vol.~XX, No.~X, August~2023}%
{Shell \MakeLowercase{\textit{et al.}}: A Sample Article Using IEEEtran.cls for IEEE Journals}


\maketitle

\begin{abstract} 
Heart sound auscultation has been applied in clinical usage for early screening of cardiovascular diseases. Due to the high demand for auscultation expertise, automatic auscultation can help with auxiliary diagnosis and reduce the burden of training professional clinicians. Nevertheless, there is a limit to classic machine learning's performance improvement in the era of big data. Deep learning has outperformed classic machine learning in many research fields, as it employs more complex model architectures with a stronger capability of extracting effective representations. Moreover, it has been successfully applied to heart sound analysis in the past years. As most review works about heart 
%BS: you mix "heart sound analysis" and "heart sounds analysis" - I decided for the first :)
sound analysis were carried out before 2017, 
%BS: changed "our" to:
the present survey is the first to work on a comprehensive overview to summarise papers on heart sound analysis with deep learning published in 2017--2022. This work introduces both classic machine learning and deep learning for comparison, and further offer insights about the advances and future research directions in deep learning for heart sound analysis. Our repository is publicly available at \url{https://github.com/zhaoren91/awesome-heart-sound-analysis}.
\end{abstract}

%~\cite{clifford2017recent}


\section{Introduction}
\label{sec:introduction}
% heart disease

% importance of auscultation
\IEEEPARstart{C}{ardiac} auscultation, \ie listening to and interpreting the heart sound, is an indispensable and critical part of the clinical examination of the patient~\cite{alam2010cardiac}. As a low-cost and non-invasive examination, cardiac auscultation is invaluable for 
%BS: sounded grammatically wrong, but I also do not understand "presence of heart" - what do you mean? Please check if ok :)
detecting a heart disease
and providing an estimate of its severity, evolution, and prognosis~\cite{hanna2002history}. Accurate cardiac auscultation may determine whether a more expensive and throughout examination should be conducted~\cite{hanna2002history}. 
Nevertheless, due to difficulties in diagnosing diastolic murmurs, the overall 
%BS: deleted "the"
sensitivity of cardiac auscultation is poor (\ie ranging from 0.21 to 1.00)~\cite{alam2010cardiac}. 
%Moreover, cardiac auscultation \re{heavily} depends on physician's \re{interpreting} skills which \re{tend to wane overtime}~\cite{noor2013heart}. 
Poor cardiac auscultation skills may either overlook significant pathology, causing deteriorating condition, or overdiagnose pathology, leading to inappropriate referral for expensive echocardiography~\cite{alam2010cardiac}.

To tackle the above problems of cardiac auscultation, classic machine learning (ML) has been widely used for automated heart sound analysis, including denoising, segmentation, and classification. For instance, support vector machines (SVMs) have been employed to detect noisy audio clips~\cite{tang2021quality}, and hidden Markov models (HMMs) have been used for heart sound segmentation~\cite{rabiner1989tutorial}. Classifiers like SVMs and decision trees have been applied to heart sound classification~\cite{whitaker2017combining,langley2017heart}. They often take hand-crafted acoustic features as the input, however, human knowledge is required for manually selecting features. Additionally, classic ML tends to excel in small-scale data, whereas its performance on large-scale datasets has remained a bottleneck.

More recently, deep learning (DL) has demonstrated its higher capability in analysing heart sounds than classic ML~\cite{ren2018heartsound}. DL models typically accept raw audio signals or time-frequency representations as the inputs~\cite{chowdhury2020time,xiao2020heart}, thereby improving the efficiency by skipping the need for selecting hand-crafted acoustic features. Complex structures of DL models also enhance their ability to learn abstract representations from large-scale datasets.

\subsection{Differences between this survey and its precursors}
There are several review studies on heart sound analysis (see~\autoref{tab:surveys}). Hand-crafted feature extraction and basic ML models (\eg SVMs and shallow artificial neural networks) were discussed in~\cite{bhoi2015multidimensional,chakrabarti2015phonocardiogram,nabih2017review,ghosh2020heart,dwivedi2018algorithms}.
DL has been used since 2016 in the PhysioNet challenge~\cite{clifford2017recent}. 
Only the studies in~\cite{clifford2017recent,chen2021deep} discussed heart sound analysis, including denoising, segmentation, and classification, along with DL approaches for partial analysis tasks. Additionally, the study in~\cite{majhi2020application} explored DL for heart sound classification. Nevertheless, heart sound segmentation with DL was not included in~\cite{clifford2017recent,chen2021deep,majhi2020application}.
This survey fills the gap in the existing reviews, as few studies have provided a comprehensive review of DL methods for heart sound analysis since 2017.
Furthermore, the state-of-the-art approaches regarding the interpretability of DL models will be summarised and discussed. This work will also summarise multiple heart sound databases, discuss the potential research problems, and outline future research directions to help promote relevant research studies.

\subsection{Challenges in Heart Sound Analysis}
Although many ML and DL methods have been applied to heart sound analysis, this research field still faces many technical challenges, including denoising, segmentation, classification, and DL model explanation.

The first challenge is denoising, which aims to remove noise from heart sounds. Recording environments can be noisy with environmental noise and speech, making denoising an essential pre-processing procedure to improve the audio quality for better segmentation and classification performance. 
%Other pre-processing procedures, such as \re{removing} lung sounds~\cite{munoz2021parallel}, are not introduced as they \re{primarily involve} signal processing methods.
%~\cite{munoz2021parallel,tsai2020blind,grooby2021new}

The second challenge is segmentation that splits a heart sound signal into multiple parts, \ie cardiac cycles or smaller segments (S1, systole, S2, and diastole). Heart sound segmentation is often a pre-processing step for classification. 

The third challenge is classification, which predicts the severity level of cardiovascular diseases or heart abnormalities from heart sounds. Heart sound classification is helpful for early screening of heart diseases in primary care.

The final one is explaining DL models for heart sound analysis. DL models, with their complex structures, often appear as black boxes to humans, despite their promising performance in heart sound analysis. In the sensitive domain of digital health, explainable DL models are crucial for clinicians to provide timely and appropriate therapies for patients. Correspondingly, trust from clinicians and patients can promote real-life application of explainable DL models.

\begin{table}[t]
  \centering
 \caption{A comparison of existing surveys on heart sound analysis. Den.: denoising, Seg.: segmentation, Cla.: classification, Int.: interpretation, mo: mentioned only.}
 \vspace{-5pt}
  \label{tab:surveys}
   \footnotesize
%  \resizebox{1.0\linewidth}{!}{
\scalebox{0.9}{
  \begin{tabular}{lcccccc}
    \toprule
   Surveys &Year& DL & Den. & Seg. & Cla. & Int.  \\
\midrule
Bhoi et al.~\cite{bhoi2015multidimensional} & by 2012& \xmark & mo & \cmark & \cmark & \xmark \\
Chakrabarti et al.~\cite{chakrabarti2015phonocardiogram} &by 2013& \xmark & \cmark & \cmark&\cmark &\xmark \\
Nabih-Ali et al.~\cite{nabih2017review} &2004-2016& \xmark &\cmark &\cmark &\cmark &\xmark \\
Clifford et al.~\cite{clifford2017recent} &2016-2017& \cmark & \cmark & \cmark & \cmark & \xmark \\
Ghosh et al.~\cite{ghosh2020heart} &by 2018& \xmark & \cmark & \cmark & \cmark & \xmark \\
Majhi et al.~\cite{majhi2020application} &2008-2018&\cmark& \xmark &\xmark&\cmark&\xmark\\
Dwivedi et al.~\cite{dwivedi2018algorithms} &1963-2018&\xmark&\xmark&\cmark&\cmark&\xmark \\
Chen et al.~\cite{chen2021deep} &2016-2020& \cmark & mo & mo & \cmark &\xmark \\
\textbf{This survey} &2017-2022& \cmark &\cmark &\cmark &\cmark &\cmark\\
    \bottomrule
  \end{tabular}}
%  }
\vspace{-10pt}
\end{table}

%BS: Changed "our" to "this"

\subsection{Contributions of this survey}
The survey has the following contributions.

\emph{The first comprehensive survey in heart sound analysis with DL.} In addition to summarising ML techniques for heart sound denoising, segmentation, and classification, this work reviews advanced DL topologies for heart sound analysis, especially segmentation, classification, and interpretation. 

\emph{Summarisation of resources.} This work summarises publicly available datasets for heart sound analysis, particularly for classification. Additionally, it provides a collection of open-sourced DL algorithms for heart sound classification.

\emph{Future research directions.} This survey discusses the limitation of current DL methods and points out potential future research topics in this area. It also discusses the importance of explainable DL models for heart sound classification, current advances, and future directions in explainable AI.

\section{Background}
\label{sec:background}

\subsection{Heart Sounds}
\begin{figure}
    \centering
    \includegraphics[width=0.9\linewidth]{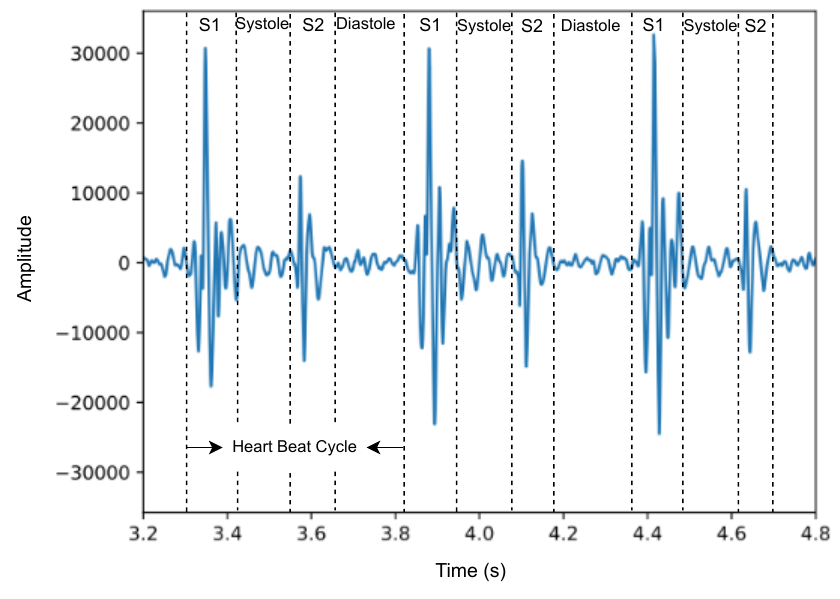}
    \vspace{-10pt}
    \caption{The PCG recording of a \emph{normal} heart sound from the PhysioNet/CinC Database~\cite{liu2016open}. Frames in the middle with four states (\ie S1, systole, S2, and diastole) are depicted.}
    \vspace{-10pt}
    \label{fig:segmentation}
\end{figure}

In a human's cardiac system, a normal cardiac cycle contains the first heart sound S1 and the second heart sound S2. S1 occurs with the closure of the mitral and tricuspid valves at the start of the systole phase, while S2 is caused by the closure of the aortic and pulmonary valves between the systole and diastole phases~\cite{bourouhou2020heart} (See Fig.~\ref{fig:segmentation}). Additionally, extra heart sounds, \ie the third heart sound S3 and the fourth heart sound S4, can occur in both normal and pathological conditions~\cite{varghees2017effective}. Both S3 and S4 manifest during the diastole phase. 
Specifically, S3 occurs after S2, resulting from the rapid filling of the ventricles, while S4 occurs before S1 (\ie at the end of diastole) during the ventricle's late filling due to atrial contraction~\cite{bourouhou2020heart,chowdhury2020time,walker1990clinical}.
The frequency range of S1 and S2 is 20--200\,Hz, while that of S3 and S4 ranges between 15--65\,Hz~\cite{naseri2013detection}.
The two types of extra sounds, S3 and S4, may indicate diseases: S3 could be a sign of heart failure~\cite{ahlstrom2006processing}, and a pathologic S4 is commonly caused by conditions that can result in ventricular hypertrophy~\cite{walker1990clinical}.

Additionally, a murmur could indicate defective valves or an orifice in the septal wall~\cite{ahlstrom2006processing}.
Murmurs, caused by turbulent blood flow in the heart system, are identified as abnormal sounds, and are crucial for diagnosing cardiovascular diseases~\cite{noor2013heart}. Murmurs often constitute the primary basis for diagnosing valvular heart disease~\cite{Gerbarg1963ComputerAO}. Clinically, murmurs consist of two types: systolic murmurs and diastolic murmurs. Aortic stenosis, mitral regurgitation, and tricuspid regurgitation occur during systole, while mitral stenosis and tricuspid stenosis occur during diastole~\cite{noor2013heart}.

% https://docs.google.com/drawings/d/1pqSn-OiAkDtXwgOgNu29bBr-N8nQunu4Z1AwSJSu56k/edit?usp=sharing
% open via diagrams.net in google drive
\begin{figure*}[t]
    \centering
    \vspace{-10pt}
    \includegraphics[ width=.7\linewidth]{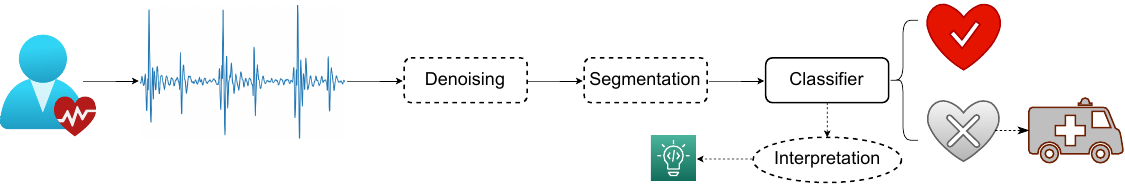}
    \vspace{-10pt}
    \caption{The framework of heart sound analysis involves denoising and segmentation, followed by training a classifier to produce the predictions and interpretations for clinicians and patients. The \ding{51} indicates a normal prediction, while the \ding{55} is an abnormal one. The dashes `-\,-' denote optional procedures.}
    \vspace{-10pt}
    \label{fig:framework}
\end{figure*}

\subsection{Diagnosis of Cardiovascular Diseases}
Nowadays, several non-invasive diagnostic tools are available for cardiovascular diseases. Particularly, medical imaging tools are capable of visualising the cardiovascular system. For instance, the echocardiogram (echo) utilises ultrasound scans to create a moving picture of the heart, offering insights into its size, shape, structure, and function~\cite{leng2015electronic}. Cardiac computed tomography (CT) uses X-rays to create detailed images of the heart and its blood vessels~\cite{leng2015electronic}. For assessment of the cardiovascular system's function and structure, cardiac magnetic resonance imaging (CMRI) creates both still and moving pictures of the heart and major blood vessels~\cite{leng2015electronic}. However, these imaging instruments are expensive and require \re{trained }medical professionals for operation, limiting their application in clinics and small- to medium-sized hospitals. 

Compared to the aforementioned diagnostic instruments, cardiac auscultation is low-cost and essential in preliminary physical examinations. Phonocardiogram (PCG) signals, recorded with a phonocardiograph, have proven to be valuable in pediatric cardiology, adult cardiology, and internal medicine~\cite{delgado2009digital}. Recent advances in electronic stethoscopes facilitated computer-aided auscultation by integrating sensor design, signal processing, and ML techniques~\cite{delgado2009digital}. The low-cost and portable nature of electronic stethoscopes makes it feasible to apply computer-aided auscultation to primary care and remote/home healthcare settings. 

Apart from PCG, Electrocardiogram (ECG), 
%BS: added (ECG is not the wave, but showing/sensing it!)
which senses the P-QRS-T wave to depict the electrical activity of the heart~\cite{martis2014current}, is an inexpensive and commonly used tool for screening heart diseases. 
ECG and PCG are highly interrelated as they are concurrent phenomena during heart activities~\cite{al2014simulation}. In an ECG signal, the P wave represents activation of the atria, followed by QRS complex resulting from ventricular excitation~\cite{al2014simulation}. The ventricles then relax back to the electrical resting state, and the T wave shows the ventricular repolarization~\cite{al2014simulation}. During this procedure, S1 occurs when the ventricles contract and the atrioventricular valves close; S2 happens when the ventricles relax and semilunar valves close~\cite{al2014simulation}. Both ECG and PCG have been used for heart abnormality detection~\cite{shekatkar2017detecting,ari2010detection}.
In~\cite{ajitkumar2021heart}, ML models with both ECG and PCG as inputs outperformed models with only one type of signal for heart abnormality detection.
%In~\cite{ajitkumar2021heart,chakir2020recognition}
Compared with PCG, ECG has difficulty in detecting structural abnormalities in heart valves and defects characterised by heart murmurs~\cite{molcer2010examination}. In this context, analysing PCG is complementary to ECG analysis in diagnosis.
%shows the potential to \re{aid} \yc{in the} diagnosis of diseases \re{that} cannot be detected from ECG. 

In Fig.~\ref{fig:framework}, heart sounds are processed by denoising, segmentation, and classification, and then clinicians and patients receive the predictions and interpretations in primary care. In real life, patients with heart sounds predicted as abnormal are recommended to do further professional medical examinations for accurate diagnosis.

\section{Heart Sound Analysis Tasks}
\label{sec:problems}

% https://docs.google.com/drawings/d/1g1P_mr-rju4ASHTHompUDx7UQFEtDxahcP6BFPB_odQ/edit
\begin{figure*}
    \centering
    \vspace{-10pt}
    \includegraphics[trim={0cm 0cm 0cm 0cm}, clip, width={.75\linewidth}]{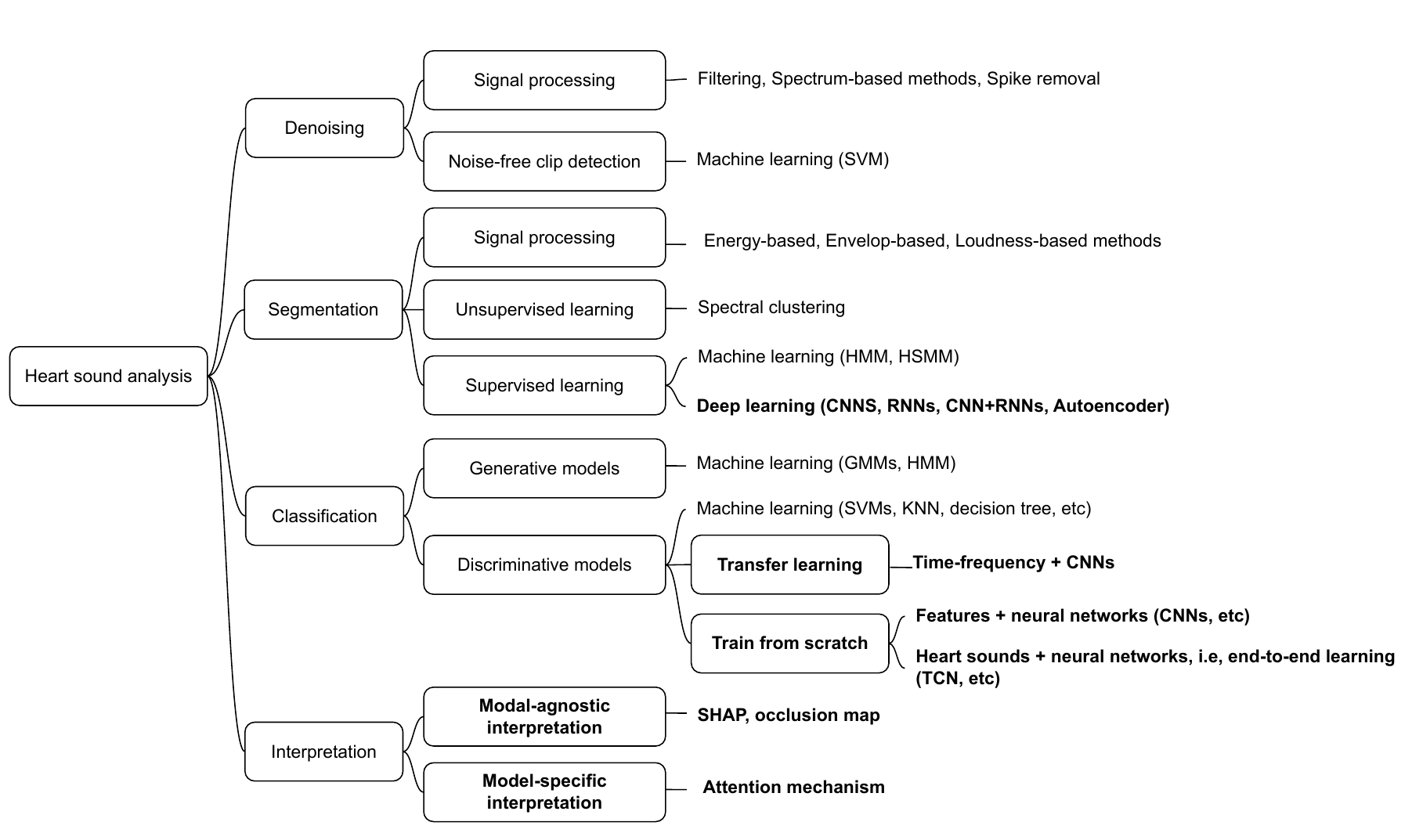}
    \vspace{-10pt}
    \caption{Categorisation of methods for heart sound analysis. Bold texts are DL approaches.}
    \vspace{-10pt}
    \label{fig:categorise}
\end{figure*}

This section describes the tasks and summarises classic ML techniques for each problem, as shown in Fig.~\ref{fig:categorise}.

\subsection{Denoising}
Generally, recorded heart sounds consist of many kinds of noises~\cite{tsao2019robust}, including white noise and other sounds presented in the recording environments, \eg human speech. Noise may impair the segmentation and classification performance of heart sounds\cite{tsao2019robust}. In this regard, numerous studies have explored denoising methods for better performance in heart sound segmentation and classification tasks. 

\sstitle{Filters} 
As a preprocessing procedure of heart sound classification, many denoising approaches employed signal filters to remove noise from heart sounds~\cite{chowdhury2020time,deperlioglu2019classification,deperliuglu2018classification}. Highpass filters have been used to eliminate low-frequency noise~\cite{alaskar2019implementation,khan2018automated}. With the capability of mitigating both high- and low-frequency noises, bandpass filters are more often used for heart sound denoising~\cite{yadav2018automatic,hu2020automatic,ghosh2019automated}. 
Butterworth bandpass filters have been successfully employed in many studies~\cite{ahmad2021automatic,singh2020short,noman2019markov,noman2019short,dastagir2021computer,bourouhou2020heart,meintjes2018fundamental,abduh2020classification}. 
The cutoff frequency of a Butterworth bandpass filter is set with a low frequency for filtering out noise with very low frequencies and a high frequency for filtering out high-frequency noises. A range of Butterworth bandpass filters with various orders have been applied with different cutoff frequency settings. For instance, a 4-th order Butterworth filter was set with a cutoff frequency of 25-400\,Hz in~\cite{singh2019classification}, and a 5-th order Butterworth filter was designed to have a cutoff frequency of 25-500\,Hz in~\cite{li2021heart} and 25-250\,Hz in~\cite{ibrahim2021comparative}. A 6-th order Butterworth filter was designed with a cutoff frequency of 50–950\,Hz~\cite{zhang2017heart,zhang2017heart1} and 30-900\,Hz in~\cite{banerjee2020multi}. 
Additionally, several other filters were also used for denoising heart sounds, such as Savitzky–Golay filter~\cite{wu2019applying,krishnan2020automated},
Chebyshev low-pass filter~\cite{zeng2021new,shervegar2018heart}, and Notch filter~\cite{al2020framework}.

\sstitle{Spectrum-based denoising}
To remove noise, spectrograms were simply selected with a threshold of -30, -45, -60, or -75\,dB in~\cite{humayun2018ensemble}. However, it is time-consuming to search for a suitable threshold among different heart sounds. A more flexible method, spectral subtraction~\cite{boll1979suppression}, was used to estimate the noise and remove it from heart sounds~\cite{abduh2020classification}.

\sstitle{Spike removal}
Frictional spike is a redundant part of the amplitude of a heart sound. In several studies~\cite{singh2019classification,noman2019markov}, frictional spikes have been detected and eliminated (\ie replaced by zeros) during pre-processing of heart sounds.

\sstitle{Selection of noise-free segments}
Apart from removing noise from heart sounds, the usage of noise-free heart sound segments has been regarded helpful for heart sound analysis.
Wavelet entropy was used as a feature to evaluate noise in heart sound segments~\cite{langley2017heart}, as clean heart sounds have relatively higher wavelet entropy than noisy heart sounds.
Empirical wavelet transform was used to separate heart sounds, murmurs, low-frequency artifacts, and high-frequency noises in another study~\cite{varghees2017effective}. 
Additionally, classic ML was also used for detecting noise-free heart sound segments. In~\cite{tang2021quality}, SVMs were applied to classify the quality of heart sound signals into binary classes (\ie `unacceptable' and `acceptable') or three classes (\ie `unacceptable', `good', `excellent').

Among the above denoising methods, filters are often used to filter out noise with defined frequency bands. Filtering, as a basic denoising approach, can be combined with other denoising methods to further improve the audio quality~\cite{abduh2020classification,singh2019classification}.
Spectral subtraction, which estimates the noise power from the frequencies outside the heart sound frequency range~\cite{abduh2020classification}, is more flexible than filters. Differently, spike removal focuses on removing the spikes rather than removing the audio components with specific frequency bands. The selection of noise-free segments is more complex than the signal-processing-based methods mentioned above, but it is more effective in automatically selecting audio segments with acceptable qualities before applying other potential denoising methods.

%%%%%%%%%%%%%%%%%%%%%%%%%%%%%%%%%%%%%%%%%%%%%%
\subsection{Segmentation}

Heart sound segmentation aims to split an audio sample into a set of smaller audio segments, which could be equal to or shorter than a complete cardiac cycle~\cite{noman2019short,baydoun2020analysis,upretee2019accurate}. 
The segments shorter than a cardiac cycle could include S1, systole, S2, and diastole, as indicated in Fig.~\ref{fig:segmentation}.  

\sstitle{Energy-based segmentation}
As heart sounds at different states have various energies, signal energy has been used for localising S1 and S2 peaks~\cite{low2018classification,deperlioglu2019classification,deperliuglu2018classification,deperlioglu2021heart}. 
Based on frequency information (\eg Wavelet Transform (WT)) of heart sounds, energy peaks of wavelet coefficients were detected for localising S1 and S2 in~\cite{eslamizadeh2017heart}. 

%BS: envelop --> envelope (everywhere!)
\sstitle{Envelope-based segmentation}
Apart from energy, heart sound segmentation can be achieved based on envelopes~\cite{akram2018analysis,ibrahim2021comparative}. For instance, Shannon energy envelope was extracted for heart sound segmentation in~\cite{babu2018automatic}. In~\cite{chowdhury2020time}, heart sound segmentation was implemented based on the Shannon energy envelope and zero crossings of heart sounds. 
In~\cite{zhang2017heart,zhang2017heart1}, S1 of the first heart cycle was detected based on Shannon energy envelopes, and subsequent S1 heart sounds were detected using a sliding heart cycle window.

\sstitle{Loudness-based segmentation}
Loudness has proven its potential to segment heart sounds~\cite{shervegar2018heart,shervegar2017automatic}.
Specifically, spectrograms extracted from heart sounds are firstly converted into the Bark scale and smoothed using a Hanning window. At each time frame, the sensation of loudness is then calculated by the mean of the amplitudes at all frequency bands: $L(t)=\frac{\sum_{t=1}^TA(t)}{T}$, where $A(t)$ is the amplitude at the $t$-th time frame, and $T$ is the total number of time frames. Furthermore, the derivation of the loudness function is computed \re{to obtain} peaks. Therefore, systoles and diastoles can be localised as they have different time lengths.

\sstitle{Classic Machine Learning for Segmentation}
ML models have been proposed for heart sound segmentation, aiming to achieve more noise-robust results than the rule-based segmentation methods mentioned above~\cite{DAS2020103990,schmidt2010segmentation,springer2015logistic}.

\emph{Unsupervised Learning.}
Considering the limited availability of the heart sound datasets, 
the authors of~\cite{DAS2020103990} adopted an unsupervised spectral clustering technique based on Gaussian kernel similarity to \re{obtain} frame labels (\eg S1 and S2), which are further utilised to segment heart sounds. 

\emph{Supervised Learning.}
Hidden Markov models (HMMs) have been widely used for segmentation~\cite{rabiner1989tutorial}. Let us assume heart states as $S=\{s_1, s_2, s_3, s_4\}=\{S1, \mbox{systole}, S2, \mbox{diastole}\}$, and the observations $O=\{o_1, o_2, ..., o_T\}$ as raw heart sounds or acoustic features. A transmission matrix $A=\{a_{ij}\}$ denotes the probablity of a state $s_i$ at the $t$-th time frame moving to $s_j$ at the $(t+1)$-th time frame. The probability density distribution of $o_t$ to be generated by $s_i$ is $B={b_i(o_t)}=P[o_t|s_i]$, where $P$ means probability. The initial state distribution is $\pi=\{\pi_i\}$, representing the probability of state $s_i$ at the start\re{ing} time frame. With $A$, $B$, $\pi$, and $O$, an HMM model aims to optimise the state sequence. The Viterbi algorithm, often used for this purpose~\cite{schmidt2010segmentation}, is further detailed in~\cite{rabiner1989tutorial}.

To better capture the abrupt changes in PCG signals, the study in~\cite{SHUKLA2020101762} \re{used signal envelops}. The kurtosis of the envelope was then computed to extract impulse-like characteristics. Subsequently, these characteristics were passed through a zero-frequency filter to obtain pure impulse information. Along with the heart sound labels, the extracted features were fed into a hidden semi-Markov model (HSMM).
In~\cite{KAMSON2019265}, a multi-centroid-duration-based HSMM was introduced to better adapt to the variability of heart cycle durations (HCDs) in PCG recordings. HCDs were estimated at various instances of a PCG to obtain maximum possible duration values, and those nearest values were clubbed into clusters to refer to each centroid. With more accurate state duration information, the HSMM achieved better segmentation performance. Similarly, considering the inter-patient variability, the emission probability distributions to each patient were estimated through a Gaussian mixture model (GMM), and an improved HSMM was used for segmentation in an unsupervised and adaptive way~\cite{8678744}. Moreover, the expectation maximisation algorithm developed in~\cite{8367820} searched for sojourn time distribution parameters of an HSMM for each subject.
Many studies~\cite{whitaker2017combining,rubin2017recognizing,maknickas2017recognition,liu2022abnormal,khan2020automatic,dastagir2021computer,bourouhou2020heart,chen2020classification,abduh2020classification,alaskar2019implementation,han2018supervised,han2019heart,chen2019heart,nogueira2019classifying} have employed logistic regression-based HSMM (LR-HSMM)~\cite{springer2015logistic} for heart sound segmentation. LR was incorporated to predict the probability of $P[s_j|o_t]$, and $B$ was then computed with \re{Bayes'} rule. 
There are also other improved HMM methods, such as the \emph{duration-dependent HMM}~\cite{schmidt2010segmentation,duggento2021classification} considering the probability density function of the duration at each state. Another study~\cite{noman2019markov} proposed a Markov-switching model for heart sound segmentation.

The energy-based, envelop-based, and loudness-based segmentation approaches attempt to detect the corresponding feature peaks, indicating S1 and S2 heart sounds. However, these approaches are primarily applied to high-quality audio samples after the denoising procedure. Classic ML methods have proven more efficient and precise in segmenting noise-contaminated heart sounds even without denoising~\cite{DAS2020103990,schmidt2010segmentation,springer2015logistic}. One can select unsupervised or supervised learning based on whether segmentation-related labels are available.

\subsection{Classification}
The goal of automated auscultation is heart sound classification, including i) detecting abnormal heart sounds (\eg murmurs, mitral stenosis, etc.) and ii) recognising severity of cardiovascular diseases (normal/mild/moderate). 

\sstitle{Feature Engineering}
Feature extraction is often performed before training a classifier. Low-level descriptors (LLDs) and functionals are typically extracted as acoustic features. LLDs represent segmental features obtained from short-time segment analysis (see Table~\ref{tab:classification_feature}), while functionals are supra-segmental feature vectors derived from LLDs. Functionals generally refer to statistical features such as mean, maximum, standard deviation, and others. 

The LLDs used for heart sound classification are listed in \autoref{tab:classification_feature}. 
In addition to time-domain LLDs, frequency-domain LLDs have been widely used for heart sound classification. Frequency-domain LLDs include (Mel-scaled) spectral features and wavelet features. There are also existing feature sets used for heart sound classification, such as the \textsc{ComParE} feature set~\cite{schuller2018compare} and the eGeMAPS feature set~\cite{eyben2015geneva}.  

In addition to hand-crafted features, more recent studies~\cite{tuncer2021application,amiriparian2018deep} have explored deep representations (see~\autoref{tab:classification_feature}). 
Given the strong capabilities of DL models in extracting abstract features, deep representations have the potential to enhance the performance of hand-crafted features.

\begin{table*}[]
    \centering
    \caption{Hand-crafted features and deep representations for heart sound classification.
    }
    \vspace{-8pt}
    \label{tab:classification_feature}  
    \begin{minipage}{\textwidth} % so footnote will appear
    \renewcommand*\footnoterule{}
    \centering
    \scalebox{0.81}{
    \begin{tabular}{p{1.5cm}p{6cm}p{7.3cm}p{4.8cm}}
    \toprule
         Group& Features & Description & Reference \\
\hline
       Time- &Envolope& Envelope of a signal & \cite{singh2019classification1,varghees2017effective}\\
         domain& Amplitude &Amplitude of a signal &\cite{eslamizadeh2017heart}\\
         &Energy &Energy of a signal &\cite{ibrahim2021comparative,khan2018automated,deperlioglu2021heart}\\
         &Entropy &Signal entropy&\cite{soares2020autonomous,khan2018automated}\\
         &Loudness&Perception of sound magnitude&\cite{shervegar2018heart}\\
         &Peak amplitude&Amplitude of peaks &\cite{dastagir2021computer}\\      
\hline 
         Spectral &Spectral amplitude & Fourier transform  &\cite{langley2017heart,khaled2022analysis,ibrahim2021comparative,arora2019heart,soares2020autonomous,yadav2018automatic,khan2018automated,sotaquira2018phonocardiogram,noman2019markov}\\
         &Dominant frequency value &Frequency which leads to the maximum spectrum &\cite{khaled2022analysis,arora2019heart} \\
         &Dominant frequency ratio & Ratio of the maximun energy to the total energy&\cite{khaled2022analysis,arora2019heart}\\
         & Energy &Spectral energy&\cite{noman2019markov}\\
         &Spectral roll-off& Frequency below a percentage of the total spectral energy&\cite{khan2018automated}\\
         &Spectral centroid &Average of magnitude spectrogram at each frame &\cite{khan2018automated,sotaquira2018phonocardiogram}\\
         &Specrtal flux & Changing speed of the power spectrum &\cite{khan2018automated}\\
         &Power spectral density (PSD) &Distribution of power in spectral components &\cite{ibrahim2021comparative,singh2019classification1,sotaquira2018phonocardiogram}\\
         &Spectral entropy & Shannon entropy of PSD&\cite{khaled2022analysis,arora2019heart,soares2020autonomous,al2020framework}\\
         &Instantaneous frequency &Frequency for non-stationary signals  &\cite{alqudah2019towards}\\
        &Fractional Fourier transform entropy&Spectral entropy of the fractional Fourier transform&\cite{tan2022heart}\\
        &Spectrogram & Short-Time Fourier Transform (STFT)  &\cite{zhang2017heart,zhang2017heart1}\\
        &Cepstrum& Inverse Fourier transform on the logarithm of the spectrum&\cite{yadav2018automatic}\\         
\hline
         Mel
        &Mel-frequency & Mel-scaled frequency &\cite{dong2019heart,maknickas2017recognition} \\
         frequency&Mel-Frequency Cepstral Coefficients (MFCCs) & Discrete cosine transform of Mel-scaled spectrogram & \cite{noman2019markov,chen2022heart,khaled2022analysis,soares2020autonomous,chen2019heart,rahmandani2018cardiac,khan2020automatic,adiban2019statistical,thiyagaraja2018novel,nogueira2019classifying} \\
         &Fractional Fourier transform-based Mel-frequency& Mel-frequency from the  fractional Fourier transform&\cite{abduh2020classification}\\
\hline
         Wavelet &Wavelet transform & Frequency analysis of a signal at various scales &\cite{singh2019classification1,baydoun2020analysis,sotaquira2018phonocardiogram}\\
         &Wavelet scattering transform &``Wavelet convolution with nonlinear modulus and averaging scaling function''\footnote{https://de.mathworks.com/help/wavelet/ug/wavelet-scattering.html} (translation invariance and elastic deformation stability~\cite{li2019heart}) & \cite{mei2021classification,li2019heart} \\
         &Wavelet synchrosqueezing transform & Reassignment of wavelet coefficients &\cite{ghosh2019automated}\\
         &Tunable quality wavelet transform & ``Wavelet multiresolution analysis with a user-specified Q-factor, the ratio of the centre frequency to the bandwidth of the filters''\footnote{https://de.mathworks.com/help/wavelet/ug/tunable-q-factor-wavelet-transform.html}&\cite{sawant2021automated,zeng2021new}\\
         &Wavelet entropy&Temporal energy distribution based on wavelet coefficients & \cite{langley2017heart}\\
\hline
         Feature set&\textsc{ComParE}& Computational Paralinguistics ChallengE feature set &\cite{dong2019heart}\\
         &eGeMAPS&The extended Geneva Minimalistic Acoustic Parameter Set &\cite{dong2019heart}\\
\hline
         Deep &Graph-based features &Petersen graph pattern &\cite{tuncer2021application}\\
         representation&Sparse coefficient&Result of sparse coding&\cite{whitaker2017combining}\\
         &Autoencoder-based features& Features extracted by an autoencoder from hand-crafted features &\cite{amiriparian2018deep,humayun2018ensemble}\\
    \bottomrule
    \end{tabular}}
    \end{minipage}
    \vspace{-20pt}
\end{table*}

\begin{figure}
    \centering
    \vspace{-10pt}
    \includegraphics[width=.98\linewidth]{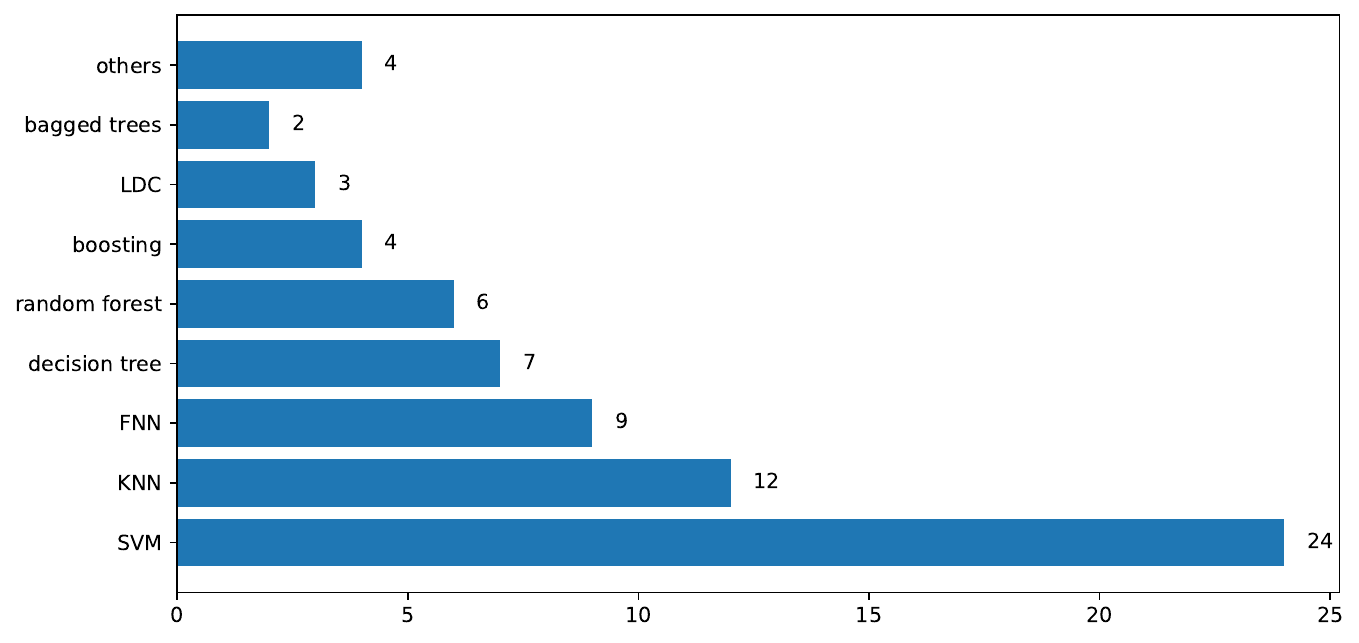}
    \vspace{-10pt}
    \caption{Statistics of the literature using discriminative ML models for heart sound classification during 2017--2022. FNN: feed-forward neural network, KNN: $k$-nearest neighbour, LDC: linear discriminant classifier.}
    \vspace{-10pt}
    \label{fig:statistic_ml}
\end{figure}

\sstitle{Classic Machine Learning for Classification}
Rule-based methods were proposed for heart sound classification in~\cite{varghees2017effective,karar2017automated}. \re{To achieve} better performance, most research studies have used classic ML for heart sound classification.

\emph{Generative models} 
aim to generate the joint probability distribution $P(X,y)$, given the features $X$ and the labels $y$. The posterior probability $P(y|X)$ is computed via \re{Bayes'} rule $P(y|X)=\frac{P(X,y)}{P(X)}=\frac{P(X|y)P(y)}{P(X)}$, where $P(X|y)$ is the likelihood probability distribution. 
The Na\"ive Bayes Classifier was widely used in heart sound classification due to its ease of use~\cite{bourouhou2020heart}.
Gaussian Mixture Models (GMMs) were used to estimate data distribution by optimising the weights of Gaussian mixture components and \re{the }mean and variance in each component~\cite{adiban2019statistical,shervegar2018heart}.
A Gaussian mixture-based HMM~\cite{noman2019markov} was employed for heart sound classification, considering the four sequential heart states. 
In~\cite{thiyagaraja2018novel}, multiple HMMs without GMMs were used for heart sound classification.

\emph{Discriminative models} are designed to directly predict the posterior probability $P(y|X)$ given $X$. Fig.~\ref{fig:statistic_ml} presents statistics of recent works from 2017 to 2022 that employ classic ML models for heart sound classification. 
SVMs have been very widely used by learning a supporting hyperplane between classes~\cite{whitaker2017combining,zhang2017heart,zhang2017heart1,tan2022heart,chen2022heart,mei2021classification,ibrahim2021comparative,tuncer2021application,dastagir2021computer,khan2020automatic,bourouhou2020heart,adiban2019statistical,upretee2019accurate,abduh2020classification,dong2019heart,amiriparian2018deep,humayun2018ensemble,ibarra2018benchmark,yadav2018automatic,chen2019heart,akram2018analysis,singh2019classification1,li2019heart,nogueira2019classifying}.
Apart from linear projection between data samples and labels, SVMs can learn to separate hyperplanes on non-linear data via non-linear kernels, such as the radial basis function. 

Furthermore, $k$-nearest neighbours (KNNs) has shown good performance in heart sound classification by classifying a data sample according to the classes of its $k$-nearest neighbours~\cite{tan2022heart,chen2022heart,ibrahim2021comparative,dastagir2021computer,khan2020automatic,singh2019classification1,alqudah2019towards,bourouhou2020heart,upretee2019accurate,abduh2020classification,ibarra2018benchmark,sofwan2019normal}.
Also, decision trees have been successfully employed in~\cite{langley2017heart,tuncer2021application,dastagir2021computer,arora2019heart,khan2020automatic,bourouhou2020heart,ibarra2018benchmark}.
One reason is that limiting the number of decision nodes can help avoid overfitting~\cite{langley2017heart}.
Additionally, the structure of a decision tree can reveal the internal logic for classification.
Bagged trees assemble multiple decision trees to create more complex model architectures for better performance~\cite{tuncer2021application,dastagir2021computer}.
Random forests further improve bagged trees with \yc{fewer} features when splitting each node~\cite{chen2022heart,arora2019heart,alqudah2019towards,ibarra2018benchmark,chen2019heart,ghosh2019automated}. 

In recent years, feed-forward neural networks (FNNs) have been applied to heart sound classification~\cite{eslamizadeh2017heart,zeng2021new,khaled2022analysis,khan2020automatic,humayun2018ensemble,khan2018automated,rahmandani2018cardiac,deperliuglu2018classification,sotaquira2018phonocardiogram}.
FNNs can automatically learn a non-linear projection between acoustic features and labels. Despite FNNs' limitation in explainability compared to classifiers like SVMs and decision trees, they show potential to achieve good performance.

There are also several other ML models, such as linear discriminant classifiers~\cite{tuncer2021application,humayun2018ensemble}, 
logistic regression~\cite{ibarra2018benchmark}, quadratic discriminant analysis~\cite{ibrahim2021comparative}, boosting methods~\cite{arora2019heart,baydoun2020analysis,chen2019heart,sawant2021automated}, and others~\cite{soares2020autonomous,al2020framework}. Finally, multiple classifiers can be further combined to enhance performance beyond what single models achieve~\cite{chen2022heart,khan2020automatic,baydoun2020analysis,abduh2020classification,singh2019classification1}

The introduced hand-crafted features and deep representations can be combined for classification. Compared to hand-crafted features, deep representations have limited explainability, making them unsuitable for interpretable decision trees. Among hand-crafted features, the listed feature groups in Table~\ref{tab:classification_feature} (\ie time-domain, spectral, Mel frequency, and Wavelet features) are complementary in representing heart sound characteristics. Therefore, they are incorporated in the \textsc{ComParE} and eGeMAPS feature sets. To enhance performance, feature selection methods can be employed to remove redundant/low-contribution features.

Generative models may require more data to model the data distribution, while discriminative models may be more susceptible to outliers. It is observed that more studies have utilised discriminative models for heart sound classification, tending to achieve good performance. However, generative models can be employed to generate additional data samples based on the learnt data distribution.

\section{State-of-the-art \zr{Deep Learning} Studies}
\label{sec:method}

DL has been successfully applied to heart sound analysis~\cite{ren2018heartsound}.
As there are few works on heart sound denoising using DL, methods for segmentation and classification with DL are introduced.

\subsection{Deep Learning for Segmentation}
%Before 2017, DL was employed for heart sound segmention~\cite{chen2016s1}. 
Various DL models~\cite{huynh2021network,duong2022deep}, categorised into convolutional neural networks (CNNs) for extracting spatial representations and recurrent neural networks (RNNs) for sequential representations, have been proposed for heart sound segmentation.

\sstitle{Convolutional neural networks}
Inspired by successful applications of deep CNNs in image segmentation, recent studies have applied deep CNNs to heart sound segmentation~\cite{8620278}. For instance, several CNN-based segmentation algorithms were proposed and compared in~\cite{8620278}, including CNNs with sequential max temporal modelling, CNNs with HMMs or HSMMs to model the probability density distribution of observations. 

\sstitle{Recurrent neural networks}
Given their capacity to leverage temporal information in sequential data, RNNs can aid in identifying the states of heart sounds.
In~\cite{8370727}, segmentation was approached as an event detection task, leading to the development of bi-directional Gated Recurrent Unit (GRU)-RNNs utilising spectrogram and envelop features.
Recognising that envelope features may inadequately capture the intrinsic duration information of heart cycles, a duration-LSTM was proposed in~\cite{CHEN2020106540}. This model integrated the duration vector into the standard LSTM cells along with envelope features, with the aim of achieving enhanced segmentation performance. Duration parameters encompass heart cycle duration and systole duration estimated from the envelope autocorrelation.
In~\cite{8749010}, the authors employed bi-directional GRU-RNNs directly for heart sound segmentation, without \re{utilising} envelopes and time-frequency based features.
Addressing the potential presence of noisy and irregular sequences in heart sound signals, an attention-based RNN framework was introduced in~\cite{8883031}. Specifically, preceding the final classification layer, a single linear layer was applied to the hidden representation returned by bi-directional LSTMs to learn the weight score of each hidden state. These weight score values are then multiplied with the hidden representation to obtain the final classification.

\sstitle{CNNs + RNNs}
In~\cite{chen2021cnn}, an end-to-end model was proposed, integrating CNNs and LSTM recurrent neural networks (RNNs) to directly learn rich and efficient features from audio waves. Furthermore, the gate structures of each LSTM unit was optimised in~\cite{8903349} for efficiency. 

\sstitle{Autoencoder}
An autoencoder comprises an encoder to map the input to hidden representations and a decoder to project the hidden features back to the input data. In~\cite{mishra2018characterization}, a stacked autoencoder was proposed for identifying S1 and S2 sounds, and the trained model outperformed deep belief neural networks as well as classic ML models like SVMs.

\subsection{Deep Learning for Classification}
\label{sec:dep_class}

\begin{figure}
    \centering
    \vspace{-10pt}
    \includegraphics[width=\linewidth]{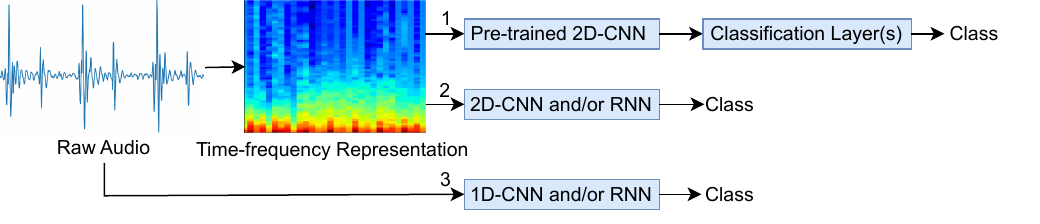}
    \vspace{-10pt}
    \caption{Pipeline of DL models working on heart sounds. ``1'' indicates transfer learning; ``2'' illustrates deep learning on the time-frequency representation; ``3'' depicts end-to-end learning. The three branches can either operate in parallel or be assembled at the feature or decision level. Additionally, DL can be utilised for processing features other than raw audio signals and time-frequency representations. 
    }
    \vspace{-10pt}
    \label{fig:dl_classification}
\end{figure}

DL employs complex models to learn effective representations directly from raw heart sounds or simple time-frequency representations. 
In addition to the pipeline shown in Fig.~\ref{fig:dl_classification}, this section summarises the advances of DL methods for heart sound classification as follows.

\sstitle{Deep learning on time-frequency representations}
Given the challenge of extracting effective representations from raw heart sound signals, 2D time-frequency representations have been widely employed as input for 2D CNNs in heart sound classification~\cite{rubin2017recognizing,maknickas2017recognition,wu2019applying,noman2019short,takezaki2021construction,duggento2021classification,chen2020classification,cheng2019design,alqudah2020classification,banerjee2020multi,han2018supervised,meintjes2018fundamental,wibawa2018abnormal,low2018classification,deperlioglu2019classification,ranipa2021multimodal}. 
In~\cite{balamurugan2019reshnet}, spectrograms extracted by STFT from heart sounds were fed into ResNet for abormal heart sounds detection.
Multi-domain features were considered more comprehensive in reflecting the characteristics of all heart sound classes.
In~\cite{wu2019applying}, spectrograms, Mel spectrograms and MFCCs were employed, and the final predictions were obtained through ensemble learning. 

In addition to CNNs, which are effective for extracting spatial features, RNNs excel at capturing temporal features. LSTM-RNNs have been applied to process discrete wavelet transforms and MFCCs for heart sound classification~\cite{ahmad2021automatic,khan2020automatic}. Deng et al.~\cite{deng2020heart} employed convolutional recurrent neural networks (CRNNs), combining CNNs and RNNs, for heart sound classification. CRNNs were also used for detecting murmurs in~\cite{wang2020automatic}. In another study, LSTMs were combined with CNNs for abnormal heart sound detection in~\cite{pandya2022infusedheart}.
Additionally, alternative classifiers have been used for heart sound classification, including a stacked sparse autoencoder deep neural network~\cite{abduh2019classification} and a semi-non-negative matrix factorisation classifier~\cite{han2019heart}.

\sstitle{Deep learning on other features}
In addition to the excellent works mentioned above on DL applied to time-frequency representations, other features extracted from heart sounds have also been utilised\re{, including time-domain features, and 1D/2D frequency-based features}.

i) \emph{Time-domain features}.
Similar to features used in classic ML approaches, 1D time-domain features can be also fed into DL models for heart sound classification. For instance, the instant energy of heart sounds was extracted as the input for stacked auto-encoder networks in~\cite{deperlioglu2021heart}. Additionally, multiple statistical features (\eg mean, median, and variance) were extracted from all 75\,ms segments in each complete heart sound clip and fed into a bidirectional LSTM (BiLSTM)-RNN model for classification in~\cite{fakhry2022comparison}.

ii) \emph{1D frequency-based features}. 
Either 1D CNNs or feed-forward DNN models can be used to process 1D features. In~\cite{ranipa2021multimodal}, general frequency features and Mel-domain features were fed into 1D CNNs, and then multiple CNNs were assembled for the final prediction. 
Mel spectrograms and MFCC were employed to extract additional features, serving as the input for a 5-layer feed-forward DNN model in~\cite{chowdhury2020time}. 

iii) \emph{2D frequency-based features}.
Herein, 2D frequency-based features are listed to include (a) multiple 1D frequency features directly extracted from audio segments rather than window functions in the STFT domain and (b) features computed from time-frequency features.
Qian et al.\ utilised wavelets to calculate wavelet energy features from a set of short acoustic segments and further used GRU-RNNs as the classifier~\cite{qian2019deep}.
Dong et al.\ extracted log Mel features and corresponding functionals from heart sound segments and implemented classification LSTM-RNNs and GRU-RNNs~\cite{dong2019heart}. In their experiments, log Mel features performed better than MFCCs and other LLDs~\cite{dong2019heart}.
Zhang et al.\ extracted temporal quasi-periodic features computed by an average magnitude difference function from spectrograms and applied LSTM-RNNs to explore the dependency relation within the features~\cite{zhang2019abnormal}. 
Additionally, a denoising auto-encoder was employed to extract deep representations from spectrograms as the input of the classifier of 1D CNNs in~\cite{li2019feature}.

\sstitle{End-to-end learning}
In recent years, as the selection of time-frequency representations and other features still requires human efforts, there has been a growing trend towards using end-to-end networks to learn representations directly from heart sounds.
Various 1D CNN architectures based on raw heart sound signals have been proposed and applied to the task of heart sound  classification~\cite{xiao2020heart,noman2019short,krishnan2020automated,hu2020automatic,avanzato2020heart}.
Furthermore, Liu et al.\ introduced a temporal convolutional network (TCN) that exhibited high sensitivity for heart sound classification~\cite{liu2022abnormal}, as a TCN benefiting from dilated and casual convolutions is more suitable for sequential data than typical CNNs. Additionally, a 1D CNN model consisting of residual blocks was developed for classifying heart sounds~\cite{oh2020classification}.
Moreover, GRU-RNNs were used to process raw heart sound signals for the screening of heart failure~\cite{gao2020gated}.

Several studies have also highlighted the capability of CNNs and RNNs in learning frequency-domain and time-domain characteristics of heart sounds. For instance, Shuvo et al.\ proposed a CardioXNet model that employed representation learning followed by sequence residual learning without any preprocessing~\cite{shuvo2021cardioxnet}. In the representation learning phase, three parallel 1D CNN pathways were constructed to extract time-invariant features from heart sound signals\re{. In} the sequence residual learning phase, BiLSTM-RNNs were employed to learn sequential representation.
The study in~\cite{li2021heart} attempted to automatically learn time-frequency features. Specifically, frequency-domain features were extracted by 1D CNNs, and the time-domain characteristics were extracted by GRU-RNNs. A self-attention mechanism was further used to fuse the two types of features for the final classification.
In~\cite{humayun2020towards,humayun2018learning}, time-convolution (tConv) layers were implemented at the front end of the network for learning finite impulse response filters.

\sstitle{Transfer learning} 
Due to extremely strict regulations governing data collection in the healthcare domain, heart sound datasets are typically not as large as datasets in other areas of Computer Audition. To overcome this limitation, transfer learning has emerged as a valuable approach, employing pre-trained DL models optimised on large-scale datasets. In recent studies, pre-trained models are primarily learnt on either an image dataset (\ie ImageNet~\cite{deng2009imagenet}) and an audio dataset (\ie AudioSet~\cite{gemmeke2017audio}). Although heart sounds are presented as audio signals, a different data type from images, DL models trained on ImageNet have demonstrated good performance on time-frequency representations extracted from heart sounds for heart sound classification. This is attributed to the fact that the large-scale ImageNet dataset improves the generalisation of DL models, and the time-frequency representaions can be fed into pre-trained DL models as colourful images. Typical DL models on ImageNet, such as AlexNet~\cite{krizhevsky2017imagenet} and VGG~\cite{simonyan2014very}, have been successfully repurposed for heart sound classification~\cite{singh2019classification,alaskar2019implementation,demir2019towards,ren2018heartsound}. Compared to ImageNet, AudioSet includes multiple types of acoustic signals and therefore is more closely related to heart sounds in terms of data type. In~\cite{koike2020audio}, pre-trained audio neural networks (PANNs) trained on AudioSet were used for classifying heart sounds with inputs of time-frequency representations. It was found that PANNs outperformed ImageNet-based models~\cite{koike2020audio}, including VGG, MobileNet V2~\cite{howard2017mobilenets}, ResNet~\cite{he2016deep}, and ResNeXt~\cite{xie2017aggregated}.

After extracting representations using pre-trained models, transfer learning uses various types of classifiers for classification, mainly including classic ML classifiers and feed-forward neural networks. For instance, SVMs were applied to representations extracted by AlexNet, VGG16, and VGG19 in~\cite{singh2019classification,demir2019towards,alaskar2019implementation}. Additionally, other classifiers such as KNNs were used in~\cite{singh2019classification}. In another approach~\cite{ren2018heartsound}, a pre-trained VGG model was frozen and followed by fully connected layers. 

Additionally, fine-tuning pre-trained models in transfer learning has also shown good performance for heart sound classification. For example, a fine-tuned AlexNet provided effective representations for heart sound classification~\cite{singh2019classification,alaskar2019implementation}. Similarly, PANNs were fine-tuned in~\cite{koike2020audio}. Fine-tuned models have even outperformed pre-trained models as they adapt to the data distribution of heart sound datasets. In~\cite{ren2018heartsound}, fine-tuned VGG performed better than pre-trained VGG when SVMs were used as classifiers.

\subsection{Interpretation}
Explainable DL has emerged as a crucial topic in healthcare. Developing explainable DL models can foster trust among physicians and patients \re{by providing insights into model predictions}. This section categorises interpretation methods into model-agnostic and model-specific approaches. Model-agnostic methods are independent of ML/DL model structures, whereas model-specific methods are closely tied to model architectures.

\sstitle{Model-agnostic interpretation}
The shapley additive explanations (SHAP) algorithm~\cite{lundberg2017unified} is based on Sharpley values, which indicate the feature importance in a prediction model according to the Game Theory. Sharpley values locally explain model predictions for each data sample, while SHAP can provide both local and global interpretations. The study in~\cite{wang2023exploring} used SHAP to locally explain a VGG model for heart sound classification. The findings of the study revealed that S1 and S2 heart sounds exhibited high feature importance in six types of time-frequency representations, including STFT, log Mel spectrogram, Hilbert–Huang transformation (HHT), wavelet transform (WT), MFCC, and Stockwell transform (ST). Interestingly, it was observed that low-frequency information in the time-frequency representations positively contributed to predicting normal heart sounds, while high-frequency information had a negative impact. Conversely, this trend was opposite for abnormal heart sounds~\cite{wang2023exploring}. 

S1 and S2 heart sounds were also found important for heart sound classification in~\cite{dissanayake2020robust}. The study~\cite{dissanayake2020robust} compared the SHAP algorithm with the occlusion map visualisation method for model interpretation. The occlusion map evaluates feature importance by masking partial feature regions, offering an alternative perspective on model explanation. It was found that the trained model might still correctly classify the heart sounds when the S1 heart sound was masked by the occlusion map. This observation suggests that other heart sound regions may also contribute to model predictions~\cite{dissanayake2020robust}.

\sstitle{Model-specific interpretation}
In~\cite{ren2022deep}, an attention mechanism was used to visualise the contribution of each feature unit to model predictions. The attention mechanism was applied to CNN, LSTM-RNN, and GRU-RNN models. It was found that the attention heatmap of heart sounds with the moderate/severe state can reveal irregular characteristics, while normal heart sounds with regular heartbeats showed regular feature importance along the time axis in the attention maps.
Similarly, in~\cite{wang2020automatic}, a temporal attention pooling mechanism was used to assign importance weights to each frame in systolic murmur regions. With the attention mechanism, the murmur regions exhibited high importance for murmurs detection.

\section{Published Resources and Advanced Performance}
\subsection{Published Datasets}
\label{sec:dataset}
In the past years, several heart sound databases have been collected. The following access-available databases are briefly introduced in~\autoref{tab:datasets}.

The PASCAL challenge Database~\cite{pascal-chsc-2011} was split into two sets A and B. In dataset A, 176 heart sounds (0.393\,hours) were recorded with the iStethoscope Pro iPhone app and annotated into S1 and S2 sounds for heart sound segmentation. Each heart sound in dataset A was also labelled into one of the four classes: \emph{normal}, \emph{murmur}, \emph{extra heart sound}, and \emph{artifact}. Dataset B with 656 recordings (1.194\,hours) was annotated into three classes: \emph{normal}, \emph{murmur}, and \emph{extrasystole}. 

The PhysioNet/CinC Database~\cite{liu2016open} used in the PhysioNet/CinC Challenge 2016~\cite{goldberger2000physionet} consists of multiple databases recorded from different data collectors. The publicly available training set includes five databases collected from both healthy individuals and patients. It comprises 3,240 recordings (20.216\,hours in total) from more than 764 subjects. The task was set as a three-class classification task: \emph{normal} and \emph{abnormal}, and \emph{noisy}. 

The Heart Sounds Shenzhen (HSS) corpus~\cite{dong2019heart} used in the INTERSPEECH Computaional Paralinguistic challengE (ComParE) 2018 was collected by the Shenzhen University General Hospital. The dataset consists of 845 recordings (7.047\,hours) from 170 subjects (f: 55, m: 115) using with an electronic stethoscope. Each audio sample was annotated into one of the three classes: \emph{normal}, \emph{mild}, and \emph{moderate/severe}. 

A heart sound dataset available on GitHub~\cite{son2018classification} contains 1,000 audio files (0.679\,hours in total). The audio recordings are balanced across five classes: \emph{normal}, \emph{aortic stenosis}, \emph{mitral regurgitation}, \emph{mitral stenosis}, and \emph{mitral valve prolapse}. 

The Michigan Heart sound database\footnote{https://open.umich.edu/find/open-educational-resources/medical/heart-sound-murmur-library}\footnote{https://www.med.umich.edu/lrc/psb\_open/html/repo/primer\_heartsound/\\primer\_heartsound.html} provides heart sounds from different areas and poses: the apex area when a subject is supine, the apex area for left decubitus, the aortic area when sitting, and the pulmonic area for supine. It consists of 23 heart sound recordings with a total duration of 0.413\,hours. The heart sounds were annotated into \emph{normal} and multiple \emph{pathological} states.

The CirCor DigiScope Database~\cite{oliveira2021circor}, used in the George B. Moody PhysioNet Challenge 2022~\cite{Reyna2022}, was collected from a pediatric population aged 21 years or younger. The heart sounds were recorded from one or multiple locations, including pulmonary valve, aortic valve, mitral valve, tricuspid valve, and others. The publicly available training set consists of 3,163 audio samples totaling with 20.094\,hours from 942 participants. Two classification tasks were targeted: i) \emph{normal} and \emph{abnormal}, and ii) \emph{presence of murmurs}, \emph{absence of murmurs}, and \emph{unclear cases of murmurs}.

\begin{table*}[!h]
  \centering
 \caption{Published datasets for heart sound classification. AS: aortic stenosis, MR: mitral regurgitation, MS: mitral stenosis, MVP: mitral valve prolapse. Notably, the statistics in this table \re{only }considered \re{accessible} data sets.}
 \vspace{-5pt}
  \label{tab:datasets}
   \footnotesize
\scalebox{0.75}{
  \begin{tabular}{llrrrl}
    \toprule
   Dataset & Challenge & \#Samples & Duration (h) & \#Subjects & Task \\
\midrule
\multirow{2}*{PASCAL Database~\cite{pascal-chsc-2011}} & \multirow{2}*{PASCAL Challenge~\cite{pascal-chsc-2011}}& 176 & 0.393 &unknown &Dataset A: Normal, Murmur, Extra Heart Sound, Artifact \\
&&656  &1.194 &unknown&Dataset B: Normal, Murmur, Extrasystole \\
\hline
PhysioNet/CinC Database~\cite{liu2016open} & PhysioNet/CinC Challenge 2016~\cite{goldberger2000physionet} & 3,240  & 20.216 &764+& Normal, Abnormal, Too noisy or ambiguous  \\
\hline
HSS~\cite{dong2019heart} & ComParE Challenge 2018~\cite{schuller2018compare} & 845  & 7.047&170& Normal, Mild, Moderate/Severe \\
\hline
Data on GitHub~\cite{son2018classification}&--& 1,000& 0.679& unknown& Normal, AS, MR, MS, MVP \\
\hline
Michigan Heart sound database\footnote{\url{https://open.umich.edu/find/open-educational-resources/medical/heart-sound-murmur-library}\url{https://www.med.umich.edu/lrc/psb_open/html/repo/primer_heartsound/primer_heartsound.html}} &--&23& 0.413 & unknown&Normal, Pathological \\
\hline
 CirCor DigiScope Database~\cite{oliveira2021circor} & George B. Moody PhysioNet Challenge 2022~\cite{Reyna2022} &3,163& 20.094& 942& Normal, abnormal; presence, absence, or unclear cases of murmurs\\
    \bottomrule
  \end{tabular}
}
\end{table*}

\subsection{State-of-the-art Performance}
As follows, the benchmarks in the challenges associated with the aforementioned databases are discussed. For databases without established benchmarks, the state-of-the-art performance in the challenges is analysed. Additionally, advanced research studies pertaining to databases not used in challenges are reviewed.

The PASCAL database was utilised in the Classifying Heart Sounds Challenge~\cite{pascal-chsc-2011}. The champion team in the challenge extracted hand-crafted features based on the segmented S1 and S2 sounds~\cite{gomes2012classifying}, and trained FNNs to classify the heart sounds. On Dataset A, the proposed approach achieved precision values of 0.35, 0.67, 0.18, and 0.92 for normal, murmur, extrasound, and artifact, respectively. On Dataset B, the method achieved precision values of 0.70, 0.30, and 0.67 for normal, murmur, and extrasystole, respectively. 

The PhysioNet/CinC challenge offered a benchmark using selected hand-crafted features and a logistic regression classifier~\cite{liu2016open}. The features were extracted from segmented four states, and then partial features were selected using logistic regression. The resulting sensitivity and specificity were 0.62 and 0.70, respectively. 
The highest average score achieved in the challenge was 0.86 (sensitivity: 0.94, specificity: 0.78)~\cite{potes2016ensemble}. 
The winning approach utilised features extracted from the above four states, which were then fed into a variant of an AdaBoost classifier. Additionally, heart sounds segmented into cardiac cycles and decomposed into multiple frequency bands were processed by a CNN classifier. Finally, an ensemble of the AdaBoost and the CNN classifiers were used for final predictions. 

The HSS database was released with a benchmark in the \textsc{ComParE} challenge 2018~\cite{schuller2018compare}. An unweighted average score of 0.562 was achieved by fusing the best two models among \textsc{ComParE} features + SVM, Bag-of-Words features + SVM, and auDeep features + RNN + SVM. The fusion strategy of majority voting outperformed multiple single-model methods. 

On the dataset available on GitHub~\cite{son2018classification}, a high accuracy of 0.988 in the five-class classification task was achieved by an SVM classifier using both MFCCs and wavelet transform features. The sensitivity and the specificity were 0.982 and 0.994, respectively. Furthermore, multiple heart valve diseases, including MR, MS, and AS, were distinguished from the healthy control with an accuracy of 0.9833 in~\cite{ghosh2020automated}.

Due to its smaller size compared to other databases, most approaches applied to the Michigan Heart sound database have used hand-crafted features and classic ML classifiers. For instance, the study in~\cite{kristomo2016heart} achieved an accuracy of 0.8889 using FNNs with hand-crafted features for a nine-class classification. Another study in~\cite{rahmandani2018cardiac} employed MFCCs and FNNs to correctly classify all samples, identifying 13 types of apex heart sounds.

The CirCor DigiScope Database~\cite{oliveira2021circor}, released in the George B. Moody PhysioNet Challenge 2022~\cite{Reyna2022}, facilitated multiple evaluation metrics, such as F-measure and accuracy. The highest weighted accuracy achieved on the test set~\cite{lu2022lightweight} was 0.78. This performance was attained by a CNN model using augmented Mel spectrograms as input for classifying heart murmurs.

\subsection{Published Algorithms}
Although only a few codes are publicly available in recent years, it is worth noting that the abundance of released codes in 2016 was largely attributed to the PhysioNet challenge 2016~\cite{liu2016open}. Notably, codes in 2016 are omitted from this study to focus on the most advanced studies during 2017--2022. 

In~\cite{son2018classification}, the authors provided a Matlab code\footnote{https://github.com/yaseen21khan/Classification-of-Heart-Sound-Signal-Using-Multiple-Features-} for training deep neural networks utilising multiple features, including MFCCs and features extracted through a discrete wavelet transform.
Additionally, the study presented in~\cite{humayun2020towards} implemented a Python-based CNN model with time-convolutional units, simulating finite impulse response filters\footnote{https://github.com/mhealthbuet/heartnet}. Furthermore, ResNets applied to linear and logarithmic spectrogram-image features were implemented in a Python code\footnote{https://github.com/mHealthBuet/CepsNET} shared by the authors of~\cite{azam2021heart}. Lastly, a Matlab code\footnote{https://github.com/uit-hdl/heart-sound-classification} for detecting valvular heart disease from heart sounds and echocardiograms was released in~\cite{waaler2022algorithm}.

%\section{Applications}

%\subsection{Denoising + Classification}

%\subsection{Segmentation + Classification}

%\subsection{Novel Applications and Settings}

\section{Future Research Directions and Open Issues}
\label{sec:discussion}

\subsection{Findings}
In classification tasks that primarily focus on the screening for heart diseases, segmentation is often considered as a preprocessing procedure before classification. The question of whether segmentation benefits classification remains open.

\sstitle{Segmentation + Classification}
Many studies have employed segmentation techniques or pre-existing segmentation information as a preprocessing step before the classification procedure. For instance, segmented cardiac cycles served as input for DL models in~\cite{humayun2020towards}. Clips beginning from the S1 heart sound with a fixed length of 1.6\,s were utilised for classification in~\cite{gao2020gated}. The importance of segmentation was demonstrated for abnormal heart sound detection in~\cite{dissanayake2020robust}.
Interestingly, experiments in~\cite{dissanayake2020robust} did not show a significant improvement in model performance when segmentation information was incorporated, compared to models without segmentation. This lack of improvement may be attributed to the inherent power and robustness of the models, suggesting that segmentation might be automatically handled by intermediate layers in these models. This assertion was supported by explanations provided by the SHAP algorithm~\cite{dissanayake2020robust}. 
Additionally, S1 and S2 sounds were observed to be more important compared to other clips within a heart sound segment. Therefore, segmentation is necessary either as an additional procedure for classifiers lacking robustness or as an internal procedure within more advanced classifiers.

\sstitle{No-segmentation} 
Several approaches have advocated for the use of non-segmented heart sounds, aiming to simplify automated auscultation~\cite{xiao2020heart,zeng2021new,singh2019classification,alqudah2019towards,takezaki2021construction}.
Apart from feeding complete heart sound samples into neural networks, heart sounds can be segmented into shorter clips of equal length for model training~\cite{wu2019applying}. For example, the first 5\,s of each audio sample were selected as the model input in~\cite{langley2017heart,singh2019classification1,chen2022heart,singh2020short} and segmented 5\,s clips were also used in~\cite{tan2022heart,arora2019heart,cheng2019design}. Most studies employ audio clips with lengths ranging from 2\,s to 6\,s~\cite{li2021heart,hu2020automatic,banerjee2020multi,krishnan2020automated}.

\subsection{Limitations and Outlook}
\sstitle{Hardware development}
In clinics, echocardiography involves obtaining ultrasound scans with a small probe that emits high-frequency sound waves. Physicians can diagnose conditions by observing the heart, blood vessels, and blood flows through this method\footnote{https://www.nhs.uk/conditions/echocardiogram/}. However, echocardiography requires well-trained skills for professionals, limiting its usage in primary care. Classic acoustic stethoscopes used in primary care require physicians and nurses to undergo training. Consequently, there is a high demand for electronic stethoscopes in primary care. In recent years, electronic stethoscopes have been developed to record heart sounds and transmit them to computers or mobile phones for further analysis~\cite{leng2015electronic}. Most electronic stethoscopes can only achieve basic functions such as amplifying and visualising heart sounds without providing a diagnosis. More recently, there are a few studies and hardware advancements focused on automated diagnosis. For instance, a field-programmable gate array (FPGA) was designed to classify heart sounds via an LSTM-RNN model in~\cite{jhong2020deep}. ``HD Steth with ECG''\footnote{https://www.stethoscope.com/hd-steth-with-ecg/} embedded artificial intelligence (AI) into an electronic stethoscope to detect multiple cardiac abnormalities. As outlined throughout this overview, AI shows promise in diagnosing heart sound abnormalities, thereby reducing the dependence on well-trained professionals. Devices capable of accurately diagnosing cardiac diseases will be invaluable in promoting early screening for cardiac conditions in primary care and home settings.

\sstitle{Performance improvement}
Although automated auscultation is ideally expected to replace human analysis, model performance can be a bottleneck for applying automated auscultation to clinical usage. False negative predictions can result in delayed or missed therapies and aggravated conditions. In future efforts, i) automated auscultation will be essential to achieve high performance and should account for individual differences in the context of personalised healthcare. The current research studies are mostly based on heart sounds, while many types of individual information such as demographics can affect model performance~\cite{thompson2019artificial}. 
Such individual information can be encoded as inputs for DL models. Additionally, electronic health records (EHR) information can be integrated to prompt personalisation. Heart sound analysis can be implemented based on heart sounds and relevant medical history, \re{thereby} providing targeted and timely diagnosis~\cite{zhao2021eires,thang2015evaluation,nguyen2022detecting}. A long-term dynamic analysis model is also essential for precise diagnosis, taking into account changes in medical status. ii) In terms of ML and DL, the field is currently witnessing the advent and adoption of foundation models \cite{bommasani2021opportunities} (pre-)trained on large-scale datasets. Several approaches using pre-trained models have been observed and listed in this work. However, one can expect even larger models to emerge with the potential for abilities directly related to heart sound analysis as a `downstream' task. On the other hand, the upcoming era of foundation models is expected to be marked by homogenisation, and it remains to be seen if the diversity of heart sound analysis approaches reported herein will indeed converge to a few large data-trained models over the next years \cite{bommasani2021opportunities}. iii) Furthermore, human-machine collaboration holds great promise for improving system performance~\cite{nguyen2014reconciling,nguyen2015smart,nguyen2015tag,hung2019handling} and providing more accurate diagnoses and timely treatments for patients. Human-machine classification can combine both machine predictions and human input (from crowd workers and experts) to achieve more precise diagnoses~\cite{callaghan2018mechanicalheart}. In~\cite{callaghan2018mechanicalheart}, data samples predicted with high uncertainty were sent to crowd workers for majority voting. Similarly, samples will be forwarded to an expert based on a certainty threshold derived from predictions made by the crowd.

\subsection{Interpretable, Dependable, and Actionable Deep Heart Sound Analysis} 
The explanation methods in Section~\ref{sec:method} offer local explanations that interpret DL models on  a case-by-case basis, yet lack a global capability of revealing the underlying classification rules or summarising the characteristics of each heart sound class. Explainable DL models~\cite{nguyen2023portable,nguyen2023example,nguyen2022model}, such as deep neural decision trees~\cite{yang2018deep}, hold promise for explaining the models themselves from a structure perspective. Learnt or searched data samples of prototypes, criticisms, and counterfactuals~\cite{van2021interpretable,ren2021prototype,chang2022example} can illustrate the typical characteristics of each class, enabling physicians to compare new samples with these heart sounds for improved understanding and analysis. Specifically, by analysing the patterns of criticisms, physicians can potentially reduce the number of false negatives, which is a crucial aspect in the healthcare field.
More recently, sonification has been proposed to explain DL models for enhanced human-computer interaction~\cite{schuller2021towards}. Unlike visualisation, sonification offers a novel perspective for explaining models through auditory means.

Additionally, considering the health implications, it appears crucial that AI-driven heart sound analysis exhibits the utmost dependability~\cite{pelillo2021machines}. While mechanisms exist, further adaptation to the specific field of application, if not novel algorithms, will need to be designed. Ultimately, dependability will emerge as a major driving factor for the trustworthiness of heart monitoring solutions. In everyday situations, trustworthiness is a key factor to winning users. 

Moreover, to enable DL models to be actionable in real life, data privacy has been another emerging research topic aiming at protecting users' data from leakage or external attacks~\cite{nguyen2024survey,ren2023anoverview}. 
Machine unlearning~\cite{NEURIPS2021_87f7ee4f,nguyen2022survey} and federated learning methods~\cite{9871319,chang2022robust} can help healthcare institutions better organise patients' private data in a secure manner without sacrificing diagnosis accuracy. 
Furthermore, AI attacks on heart sound analysis, such as through adversarial attacks, needs to be contemplated and dealt with.
In summary, DL models hold promise in guiding healthcare providers' actions in their daily practices to provide better care for patients. It will be essential to improve DL models not only in terms of performance but also from human-centred perspectives in future.

\subsection{Real-life Applications}
PCG signals hold promise for applications in abnormal heart sound detection, typically approached as a binary classification task. For predicting specific abnormal heart sounds, several databases, including the PASCAL database~\cite{pascal-chsc-2011}, HSS~\cite{dong2019heart}, and data available on Github~\cite{son2018classification}, have enabled multi-class classification tasks. More specific predictions of heart sounds provide more precise early screening, benefiting both patients and clinicians compared to binary classification. However, the above three databases are not as large as the PhysioNet/CinC database~\cite{liu2016open} and the CirCor DigiScope database~\cite{oliveira2021circor} (see Table~\ref{tab:datasets}). Collecting heart sound data with more detailed labels can be a further step in heart sound analysis, potentially facilitating the release of more databases for relevant research. Furthermore, algorithms such as transfer learning and multi-task learning, which involve both binary classification and multi-class classification, have the potential for achieving more detailed predictions.

In real-life scenarios, automated heart sound analysis holds promise as an early screening tool for patients. The study in~\cite{pandya2022infusedheart} presented a pipeline from heart sound recording to abnormal heart sound detection in real-life usage. Once a tool with heart sound analysis is developed, it can be used for real-time heart sound analysis on a daily basis. Herein, recorded data can be processed either online on a cloud server or offline, taking into account users' privacy issues. The study in~\cite{9871319} introduced a federated learning framework for heart sound analysis. Furthermore, users can go to clinics for further diagnosis if abnormal heart sounds are detected. This can help the users promptly seek medical attention, and also reduce the burden on clinics with long waiting lists. In clinics, physicians can benefit from the analysis model's outputs, including predictions and interpretations, to guide the use of suitable diagnosis instruments and provide effective treatments. If the tool is granted permission to monitor patients' heart status, physicians can offer timely and useful suggestions to patients. Additionally, detecting other health statuses from heart sounds shows promise in developing comprehensive healthcare instruments. For example, psychological stress was detected from simultaneous PCG and ECG signals~\cite{cheema2019psychological}.

\section{Conclusion}
This work summarised both classic machine learning and deep learning technologies for heart sound analysis from 2017 to 2022, including denoising, segmentation, classification, and interpretation. Available databases were introduced with evaluation metrics in this study. This work also listed publicly available repositories for implementing heart sound classification. Additionally, several findings and limitations of heart sound classification were analysed, and potential future works were discussed. 
This work presented a summary of the advances in heart sound analysis, provided insightful discussions, and highlighted promising research directions for the community.

\section*{Acknowledgement}
This research was funded in part by the Federal Ministry of Education and Research (BMBF), Germany under the project LeibnizKILabor with grant No.\,01DD20003, the Ministry of Science and Technology of the People's Republic of China with the STI2030-Major Projects (No.\,2021ZD0201900), the National Natural Science Foundation of China (No.\,62272044), and the Teli Young Fellow Program from the Beijing Institute of Technology, China. The research was mainly done when Zhao Ren was at the L3S Research Center, Leibniz University Hannover, Germany. The last author further acknowledges help from the Munich Data Science Institute (MDSI) and the Munich Center for Machine Learning (MCML) -- both in Munich, Germany.

%\bibliographystyle{ieeetr}
%\bibliography{cim/paper,cim/ref_h}

%\appendices

%\vfill

\end{document}